\documentclass{ws-ijmpc}

\usepackage{ulem}
\usepackage{hyperref}
\usepackage{lipsum,verbatim}
\usepackage{graphicx}
\usepackage{multirow,rotate,subfigure}
\usepackage{amscd,amsmath,amssymb}
\usepackage{booktabs}

\usepackage{enumitem}

\usepackage{adjustbox}

\usepackage{booktabs}

\usepackage{mathrsfs}
\usepackage{mathtools}

\usepackage{amscd,amssymb}
\usepackage{mathtools}

\usepackage{color, colortbl}
\definecolor{red}{rgb}{1,0,0}
\definecolor{green}{rgb}{0,1,0}
\definecolor{blue}{rgb}{0,0,1}
\definecolor{Gray}{gray}{0.9} 
\definecolor{LightCyan}{rgb}{0.88,1,1}
\definecolor{ciano}{rgb}{0,1,1}
\definecolor{magenta}{rgb}{1,0,1}
\definecolor{amarelo}{rgb}{1,1,0}
\definecolor{bananayellow}{rgb}{1.0, 0.88, 0.21}

\newcommand{\xxb}[1]{\textcolor{blue}{#1}}

\usepackage{color, colortbl}
\definecolor{babypink}{rgb}{0.96, 0.76, 0.76}
\definecolor{amarelo}{rgb}{0.91, 0.84, 0.42}  
\definecolor{aqua}{rgb}{0.0, 1.0, 1.0}  
\definecolor{iw}{rgb}{0.7,0.93,0.36}  
\definecolor{lav}{rgb}{0.96,0.73,1.0}  
\definecolor{awesome}{rgb}{1.0, 0.13, 0.32}    
\definecolor{cadetgrey}{rgb}{0.57, 0.64, 0.69} 
\definecolor{x11gray}{rgb}{0.75, 0.75, 0.75}   
\definecolor{bananayellow}{rgb}{1.0, 0.88, 0.21}

\newcommand{\iw}{\cellcolor{iw}}
\newcommand{\iy}{\cellcolor{lav}}
\newcommand{\am}{\cellcolor{amarelo}}
\newcommand{\bl}{\cellcolor{aqua}}
\newcommand{\aw}{\cellcolor{awesome}}

\newcommand{\by}{\cellcolor{bananayellow}}

\begin{document}

\markboth{Jason A.C.~Gallas}
{Carriers and inheritance in quadratic dynamics}

\catchline{}{}{}{}{}

\title{Orbital carriers and inheritance in discrete-time quadratic dynamics}

\vspace{-0.25truecm}

\author{Jason A.C.~Gallas} 

\address{
Instituto de Altos Estudos da Para\'\i ba,
  Rua Silvino Lopes 419-2502,\\   
  58039-190 Jo\~ao Pessoa, Brazil,\\
Complexity Sciences Center, 9225 Collins Ave.~Suite 1208, 
Surfside FL 33154, USA,\\
  Max-Planck-Institut f\"ur Physik komplexer Systeme,
  01187 Dresden, Germany\\
     jason.gallas@gmail.com}

\maketitle

\begin{history}
\received{2 March 2020}
\accepted{24 March 2020}
\centerline{Published 22June 2020}
\centerline{https://doi.org/10.1142/S0129183120501004}
\end{history}


\begin{abstract}
Explicit formulas for {\sl orbital carriers}
of periods $4$, $5$, and $6$ are reported for  discrete-time quadratic
dynamics. 
A systematic investigation of {\sl orbital inheritance} for
periods as high as  $k\leq 12$ is also reported.
Inheritance means that unknown orbits may be obtained
by nonlinear transformations of known orbits.
Such nested {\sl orbit within orbit stratification} shows orbits
not to be necessarily independent of each other as generally assumed.
Orbital stratification is potentially significant to rearrange
trajectories sums in trace formulas underlying modern semiclassical
interpretations of atomic physics spectra.
The stratification seems to dominate as the orbital period grows.
\keywords{Orbital carriers; Orbital inheritance;
          Quadratic dynamics; Symbolic computation.}
\end{abstract}

\ccode{PACS Nos.:
      02.70.Wz, 
      02.10.De, 
      03.65.Fd} 

\section{Introduction}

Applied problems in physics normally require solving equations of motion,
frequently expressed either as
differential equations involving continuous-time derivatives, or
discrete-time maps.
Since the advent of modern computers, solving equations of motion
essentially boils down to number crunching using special-purpose
numerical methods.
For a representative selection of methods and applications see, e.g., 
Refs.~\cite{bk1,bk2,bk3}. 

Numerical methods revealed much of what is presently known about 
the time-evolution of complex systems.
However, there are certain peculiarities that are totally out of reach to
approximate numerical methods and that have not yet been addressed
as they could.
For instance, consider cascades of periodic motions, which are among the
most prominent features found in systems governed by 
differential equations or by maps. 
Although such cascades cannot be followed analytically for differential
equations, they are accessible in systems governed by maps with algebraic
equations of motion,
particularly in one-dimensional dissipative maps\cite{bk1,bk2,bk3}.

Consider a popular class of models, namely one-dimensional maps governed
by algebraic equations of motion.
To delimit analytically their stability windows one needs to solve polynomials
containing physical parameters.
Although parameters may vary freely, conditions imposed on the stability
boundaries greatly reduce such freedom as well as the complexity of the
numerical values defining boundaries.
For example, for the paradigmatic quadratic and H\'enon maps, the polynomial
coefficients at intersections are simply given by integers or by algebraic
numbers\cite{bk1,bk2,bk3,jg95,jg19}.

The self-similar regularities recorded for cascades of periodic motions in
parameterized maps pose a natural question regarding the generic
arithmetic nature of the numbers delimiting adjacent windows of stability
as parameters are varied.
Knowledge of the arithmetical unfolding of such cascades should
provide insight into the analytical clockwork mechanism underlying this
forever repeating process.
Such information cannot be inferred from approximate computations
but could be eventually won using exact algebraic analysis.

The purpose of this paper is to report a systematic investigation of
orbital carriers and 
{\sl orbital inheritance} in discrete-time quadratic dynamics, in the
so-called partition generating limit\cite{bk1,bk2}, whose equation of
motion is $x_{t+1}=2-x_t^2$. More specifically, we extend previous
work\cite{jg19}
to include carriers for orbits with periods $4$, $5$ and $6$, and
inheritance for periods $k\leq 12$ of the map.
Such computations are quite strenuous.
The latter limit is set
by the capability of the hardware and software at our disposal to generate
and to factor large polynomials of degrees no less than 4020, with
exceedingly large numerical coefficients and discriminants.
As discussed and illustrated below,
carriers are polynomials encoding simultaneously {\sl all possible orbits
of a given period}\cite{jg19}.
Inheritance means that known periodic orbits reveal unknown orbits.
New orbits are obtained through simple nonlinear
transformations from known orbits\cite{epl,jg00,prep}.
Inherited orbits are {\sl clones} that share an arithmetic ancestry.
Arithmetic interdependencies among periodic orbits are hard, not to say
impossible, to recognize in numerical simulations, where only approximate 
numbers are considered. 

The starting point to investigate the arithmetic nature of equations of motion
is the ring $\mathbb Z$ of integers, namely solving polynomials with
integer coefficients.
Key properties which facilitate the study of polynomials with integer
coefficients are the Euclidean algorithm and
the unique factorization of integers (the `fundamental theorem of arithmetic').
Such properties no longer always hold for rings of integers of higher
algebraic number fields, involving polynomials with a good deal more
complicated coefficients and which are the framework where algebraic
equations of motions must be considered.

The first coherent discussion of
complex integers $a+ib$ with rational integral $a$ and $b$ was
presented by Gauss as far back as {1831-32, in his second paper on
biquadratic reciprocity.
Subsequently, the theory of quadratic algebraic numbers was
essentially completed during the nineteen
century by Kummer, Dirichlet, Dedekind, Hilbert and others\cite{som07}.
However, the corresponding knowledge regarding numbers as simple as
{\sl cubics and relative cubics} is by far less complete,
despite more than two centuries of work\cite{df64}.
The main difficulty comes from the well-known fact that irreducible cubics
with three real roots, the so-called {\it casus irreducibilis}, cannot
have their roots expressed in terms of real radicals.
The equations of motion discussed here are attractive in that they require
investigating towers of such cubic fields.
We consider periods $k\leq12$, and provide explicit solutions
for polynomials of degrees as high as 18 and 24, involving
nested cubic roots.

With respect to applications beyond the scope of dynamical systems,
we mention briefly that the concept of inheritance is potentially attractive
for atomic physics, where it seems to imply the interesting and
unsuspected possibility of rearranging
certain orbit-dependent contributions in cycle expansions and
semiclassical sums needed for calculating energy spectra and density of
states using, e.g., Gutzwiller's
trace formula\cite{h10,m06,b99,alt,b01,sh98,mw98,gd99}.

\section{Orbital carriers for periods $4$, $5$ and $6$}

A recent work has shown that classical equations of motion
of algebraic origin may be all conveniently extracted from
just a single mathematical object, a polynomial called an
{\sl orbital carrier}.
All possible orbits may be encoded simultaneously by a single carrier,
with individual orbits parameterized by $\sigma$, the sum of their
orbital points\cite{jg19}.
In Ref.~\cite{jg19}, such parameterization
was established for period-three orbits using standard textbook
knowledge of the theory of algebraic equations.
Essentially, one uses certain functions of the roots of the equation
of motion, the {\sl elementary symmetric functions}\cite{stan99},
which may be expressed in a general manner by means of the coefficients
of the equation of motion, without the equation itself being resolved.
This fact shifts the traditional study of orbital {\it points} to
a new level, to the study of orbital {\it equations} of motion.

Here, we extend the aforementioned orbital parameterization to include
explicit expressions for carriers of periods $4$, $5$ and $6$.
Results for periods four\cite{eg02} and six\cite{eg04}
may be obtained as particular cases of general expressions obtained
previously for the two-parameter H\'enon map,
$(x,y) \mapsto (a-x^2+by, x)$.
For arbitrary values of $a$, carriers for the quadratic map
$x_{t+1} = a-x_t^2$ are obtained setting $b=0$ in the expressions
of the H\'enon map.
For the partition generating limit discussed here, set $(a,b)=(2,0)$.
The carrier for period five is freshly obtained and is reported here
for the first time. 
Apart from the theoretical novelty of these carriers, they help to
motivate the main results below and to make them more comprehensible.

\subsection{The period four carrier}

Essentially, for a given period $k$, all period-$k$ orbits may be
encoded simultaneously by two polynomials, as described in a recent
open access paper\cite{jg19}:
A $\sigma$-parameterized polynomial $\psi_k(x)$, called the carrier,
and an auxiliary polynomial, $\mathbb S_k(\sigma)$, which fixes the values
of the parameter $\sigma$ for each individual orbit.
The parameter $\sigma$ is just the sum of the orbital points.
The degree of the polynomial $\mathbb S_k(\sigma)$ informs
the total number of possible $k$-periodic orbits in the system.
When substituted into $\psi_k(x)$, each individual root of
$\mathbb S_k(\sigma)=0$ ``projects'' $\psi_k(x)$ into the $\sigma$-selected
individual orbit.
Normally,  $\mathbb S_k(\sigma)$ is a reducible polynomial over the integers:
nonlinear factors of degree  $\partial_k$ correspond to
{\sl orbital clusters}, namely to irreducible polynomial aggregates
commingling together a total of $\partial_k$  orbits.
Linear factors correspond to non-clustered {\it single} orbits of degree $k$.

For period-four there are three possible orbits, all {\sl encoded\/}
simultaneously by the doublet:
{\small
\begin{eqnarray}
 \psi_4(x) &=&{x}^{4}-\sigma{x}^{3} + \tfrac{1}{2}({\sigma}^{2}+\sigma-8){x}^{2}
  -\tfrac{1}{6} ({\sigma}^{3}+3\,{\sigma}^{2}-20\,\sigma+2) x\cr
  &&\quad\ +\tfrac{1}{24} (\sigma-3)( {\sigma}^{3}
       +9\,{\sigma}^{2}-2\,\sigma-16) \label{o4}\\
\mathbb S_4(\sigma) &=& (\sigma+1)(\sigma^2-\sigma-4).  \label{a4}
\end{eqnarray}}
Substituting $\sigma=-1$ into $\psi_4(x)$ we obtain the orbit $o_{4,1}(x)$,
while for $(1-\sqrt{17})/2$ and $(1+\sqrt{17})/2$, roots of the quadratic
factor, we get $o_{4,2}(x)$ and $o_{4,3}(x)$, respectively:
{\small
\begin{eqnarray}
o_{4,1}(x) &=& {x}^{4}+{x}^{3}-4\,{x}^{2}-4\,x+1,\\
o_{4,2}(x) &=&{x}^{4} -\tfrac{1}{2}(1-\sqrt {17}) {x}^{3}
       - \tfrac{1}{2}(3+\sqrt {17}) {x}^{2}  - (2+\sqrt {17}) x-1,\label{o42}\\
o_{4,3}(x)  &=& {x}^{4} -\tfrac{1}{2}(1+\sqrt {17}) {x}^{3}
       - \tfrac{1}{2}(3-\sqrt {17}) {x}^{2}  - (2-\sqrt {17}) x-1.\label{o43}
\end{eqnarray}}
When multiplied together, $o_{4,2}(x)$ and $o_{4,3}(x)$ produce the orbital
cluster, or aggregate:
{\small
\begin{equation}
  c_{4,1}(x) = o_{4,2}(x)\cdot o_{4,3}(x)
            ={x}^{8}-{x}^{7}-7\,{x}^{6}+6\,{x}^{5}+15\,{x}^{4}
                       -10\,{x}^{3}-10\,{x}^{2}+4\,x+1,
\end{equation}
a cluster that may be obtained directly by eliminating  $\sigma$ between
$\psi_4(x)$ and $\sigma^2-\sigma-4$.

Note that the product of $o_{4,2}(x)$ and $o_{4,3}(x)$, which have algebraic coefficients,
resulted in a cluster with integer coefficients, a generic characteristic.
Technically, $o_{4,2}(x)$ and $o_{4,3}(x)$ are defined by {\sl relative quadratic}
equations of motion\cite{som07}.
Manifestly,  $c_{4,1}(x)$ decomposes over the field $\mathbb Q(\sqrt{17})$.
The orbit $ o_{4,1}(x)$ has always integer coefficients
and is always an exact representation for the orbit. In sharp contrast,
when projected onto the real axis, $o_{4,2}(x)$ and $o_{4,3}(x)$
will have necessarily approximate numerical coefficients.
Thus, the symmetries clearly visible between Eqs.~(\ref{o42}) and (\ref{o43})
will be totally obliterated.
This unambiguous dichotomic distinction between orbits remains valid for  other
periods and neatly displays the enhanced insight obtained by working
with exact equations of motion.

Doublets like Eqs.~(\ref{o4}) and (\ref{a4}) may be determined for arbitrary
periods. Expressions for arbitrary values of $a$ of the quadratic map and arbitrary
$(a,b)$ of the H\'enon map are available\cite{eg02,eg04}.

\subsection{The period five carrier}

For period-five there are six possible orbits, all encoded simultaneously
by the doublet:
{\small
\begin{eqnarray}
\psi_5(x) &=&( 360\,{\sigma}^{2}-360\sigma-240) {x}^{5}
-120\sigma(3{\sigma}^{2}-3\sigma-2) {x}^{4}+60({\sigma}^{2}
       +\sigma-10)(3{\sigma}^{2}-3\sigma-2) {x}^{3}\cr
     &&\quad -(60{\sigma}^{5}+90{\sigma}^{4}-1800{\sigma}^{3}+1710{\sigma}^{2}
               +1860\sigma-1200){x}^{2}\cr
     &&\quad+(15{\sigma}^{6}+45{\sigma}^{5}-735{\sigma}^{4}
               +375{\sigma}^{3}+3480{\sigma}^{2}-2700\sigma
             -1200) x\cr
     &&\quad -3{\sigma}^{7}-12{\sigma}^{6}+192{\sigma}^{5}+30{\sigma}^{4}
             -2061{\sigma}^{3} +1446{\sigma}^{2}+4248\sigma-3600,\\
\mathbb S_5(\sigma) &=&
 (\sigma-1) ({\sigma}^{2}+\sigma-8) ({\sigma}^{3}-{\sigma}^{2}-10\sigma+8).
\end{eqnarray}
Eliminating $\sigma$ between  $\psi_5(x)$ and, successively, $\sigma-1$,
${\sigma}^{2}+\sigma-8$, and ${\sigma}^{3}-{\sigma}^{2}-10\sigma+8$, we get,
apart from multiplicative constants used to eliminate denominators in
$\psi_5(x)$,
{\small
\begin{eqnarray}
o_{5,1}(x) &=& {x}^{5}-{x}^{4}-4\,{x}^{3}+3\,{x}^{2}+3\,x-1,\\
c_{5,1}(x) &=& {x}^{10}+{x}^{9}-10\,{x}^{8}-10\,{x}^{7}+34\,{x}^{6}
              +34\,{x}^{5}-43\,{x}^{4}-43\,{x}^{3}+12\,{x}^{2}+12\,x+1,\\  
c_{5,2}(x) &=& {x}^{15}-{x}^{14}-14\,{x}^{13}+13\,{x}^{12}+78\,{x}^{11}
                 -66\,{x}^{10}-220\,{x}^{9}+165\,{x}^{8}+330\,{x}^{7}\cr
   &&\quad-210\,{x}^{6}-252\,{x}^{5}+126\,{x}^{4}+84\,{x}^{3} -28\,{x}^{2}-8\,x+1.
\end{eqnarray}
The clusters factor into quintics over $\mathbb Q(\sqrt{33})$ and
$\mathbb Q\big(\sqrt[3]{-62+95\sqrt{-3}}\,\big)$, respectively, thereby
providing explicit expressions for the remaining five period-five orbits.
As before, clustered orbits involve relative quadratic and cubic equations,
with algebraic (non-integer) coefficients which,
in numerical computations cannot be determined exactly.

\subsection{The period six carrier}

For period-six there are nine possible orbits, all encoded simultaneously
by the doublet:
{\small
\begin{eqnarray}
  \psi_6(x) &=&  160 \varphi^{2}{\sigma}^{2} ({x}^{6}-\sigma\,{x}^{5})
     \ +\ 80\,{\sigma}^{2}(\sigma+4)
   ( \sigma-3)  \varphi^{2}{x}^{4}\cr
  &&-40\,\sigma\, \varphi \big( 2\,{\sigma}^{7}+{\sigma}^{6}-86\,{\sigma}^{5}+126\,{\sigma}^{4}
               +358\,{\sigma}^{3}-343\,{\sigma}^{2}-50\,\sigma+56 \big) {x}^{3}\cr
  &&+20\,{\sigma}^{2} \varphi  \big( {\sigma}^{7}+3\,{\sigma}^{6}
               -70\,{\sigma}^{5}+48\,{\sigma}^{4}
               +679\,{\sigma}^{3}-683\,{\sigma}^{2}-1218\,\sigma+1048 \big) {x}^{2}\cr
  &&-4\,\sigma\, \varphi  \big( {\sigma}^{9}+6 \,{\sigma}^{8}-91\,{\sigma}^{7}-78\,{\sigma}^{6}
               +1693\,{\sigma}^{5}-976\,{\sigma}^{4}-6911\,{\sigma}^{3}
               +5496\,{\sigma}^{2}+2508\,\sigma-2128 \big) x\cr
  &&\qquad+2\,{\sigma}^{14}+14\,{\sigma}^{13}-247\,{\sigma}^{12}-268\,{\sigma}^{11}
           +7984\,{\sigma}^{10}-8072\,{\sigma}^{9}
           -80966\,{\sigma}^{8}+157668\,{\sigma}^{7}\cr
  &&\qquad+184938\,{\sigma}^{6} -530694\,{\sigma}^{5}+88965\,{\sigma}^{4}
      +373032\,{\sigma}^{3}-197156\,{\sigma}^{2}-13440\,\sigma+15680, \label{o6}\\
\mathbb S_6(\sigma) &=& (\sigma+1)(\sigma-1)( {\sigma}^{3}-21\,\sigma+28 )
  ( {\sigma}^{4}+{\sigma}^{3}-24\,{\sigma}^{2}-4\,\sigma+16).  \label{aaa}           
\end{eqnarray}}\noindent
where  $\varphi\equiv 3\,{\sigma}^{3}-7\,{\sigma}^{2}-13\,\sigma+13$.
Apart from  multiplicative constants used to eliminate denominators in $\psi_6(x)$,
by selecting $\sigma=1$ and $\sigma=-1$ we get the orbits and discriminants:
\begin{eqnarray}
 o_{6,1}(x) &=& {x}^{6}-{x}^{5}-5\,{x}^{4}+4\,{x}^{3}+6\,{x}^{2}-3\,x-1,
    \qquad \Delta_{6,1} = 13^5=371293, \label{o61}\\
o_{6,2}(x) &=& {x}^{6}+{x}^{5}-6\,{x}^{4}-6\,{x}^{3}+8\,{x}^{2}+8\,x+1,
     \qquad \Delta_{6,2}= 3^3\cdot 7^5=453789. \label{o62}
\end{eqnarray}

Again, for roots of the cubic and quartic factors in Eq.~(\ref{aaa}),
the resulting coefficients in Eq.~(\ref{o6}) are more complicated
algebraic numbers, not integers.
When all orbits arising from the same $\sigma-$factor are multiplied together
one obtains a cluster, a polynomial aggregate with integer coefficients and
degree $\partial=mk$, multiple of the period $k$, where $m>1$ is an integer:
{\small
\begin{eqnarray}
c_{6,1}(x) &=&   
   {x}^{18}-18\,{x}^{16}+{x}^{15}+135\,{x}^{14}-15\,{x}^{13}-546\,{x}^{12}
             +90\,{x}^{11}+1287\,{x}^{10}-276\,{x}^{9}\cr
  &&\quad -1782\,{x}^{8}+459\,{x}^{7}+1385\,{x}^{6}-405\,
             {x}^{5}-534\,{x}^{4}+170\,{x}^{3}+72\,{x}^{2}-24\,x+1,\\
c_{6,2}(x) &=&{x}^{24}+{x}^{23}-24\,{x}^{22}-23\,{x}^{21}+252\,{x}^{20}
    +229\,{x}^{19}-1521\,{x}^{18}-1292\,{x}^{17}+5832\,{x}^{16}\cr
   && +4540\,{x}^{15}-14822\,{x}^{14}-10282\,{x}^{13}
    +25284\,{x}^{12}+15001\,{x}^{11}-28667\,{x}^{10}-13653\,{x}^{9}\cr
   && +20886\,{x}^{8}+7168\,{x}^{7}-9126\,{x}^{6}-1802\,{x}^{5}
+2085\,{x}^{4}+101\,{x}^{3}-180\,{x}^{2}+12\,x+1. 
\end{eqnarray}}
Independently from $\psi_6(x)$, the Maple driver given in Appendix A
exemplifies how to extract $o_{6,1}(x)$, $o_{6,2}(x)$, $c_{6,1}(x)$,
and $c_{6,2}(x)$ directly from the quadratic equation of motion.

{\small
\begin{table}[!tbh]
\tbl{The nine period-six orbits $o_{6,j}$ of the map $x_{t+1}=2-x_t^2$.
  Here, $\sigma_{6,j}=\sum x_j$ is the sum of the orbital points.
  The triad and quartet of $\sigma_{6,j}$ values are roots of the cubic
  and quartic factors in Eq.~(\ref{aaa}), respectively.
}
{\begin{tabular}{@{}|c||c|c|c|c|c|c||c|@{}} 
\hline
 {Orbit} & $x_1$ & $x_2$ & $x_3$ & $x_4$ & $x_5$ & $x_6$ & $\sigma_{6,j}$\\ 
\hline
$o_{6,1}$ & -1.770912051306 & -1.1361 &  0.7093 & 1.4969 & -0.2407 & 1.9421 & 1\\
\hline
$o_{6,2}$ & -1.911145611572 & -1.6523 & -0.7301 & 1.4670 & -0.1521 & 1.9769 &-1\\
\hline
$o_{6,3}$ & -1.990061550730 & -1.9605 & -1.8436 & -1.3989&  0.0431 & 1.9981 & -5.142457360\\
$o_{6,4}$ & -1.756443146740 & -1.0849 &  0.8230 &  1.3227&  0.2505 & 1.9372 &  1.491252188\\
$o_{6,5}$ & -0.912421314706 &  1.1675 &  0.6369 &  1.5944& -0.5421 & 1.7061 &  3.651205171\\
\hline  
$o_{6,6}$ & -1.990663269435 & -1.9629 & -1.8530 & -1.4336& -0.0552 & 1.9970 & -5.287613777\\ 
$o_{6,7}$ & -1.916491658218 & -1.6730 & -0.7989 &  1.3618&  0.1455 & 1.9788 & -0.902246984\\ 
$o_{6,8}$ & -1.559348708126 & -0.4314 &  1.8139 & -1.2902&  0.3354 & 1.8875 &  0.756484903\\
$o_{6,9}$ & -0.971966826485 &  1.0553 &  0.8863 &  1.2145&  0.5250 & 1.7244 &  4.433375858\\
\hline
\end{tabular}}\label{tab:tab01}
\end{table}
}


{\small
\begin{figure*}[!thb]
\centering
\includegraphics[width=0.85\textwidth,viewport=102 258 517 670,clip]{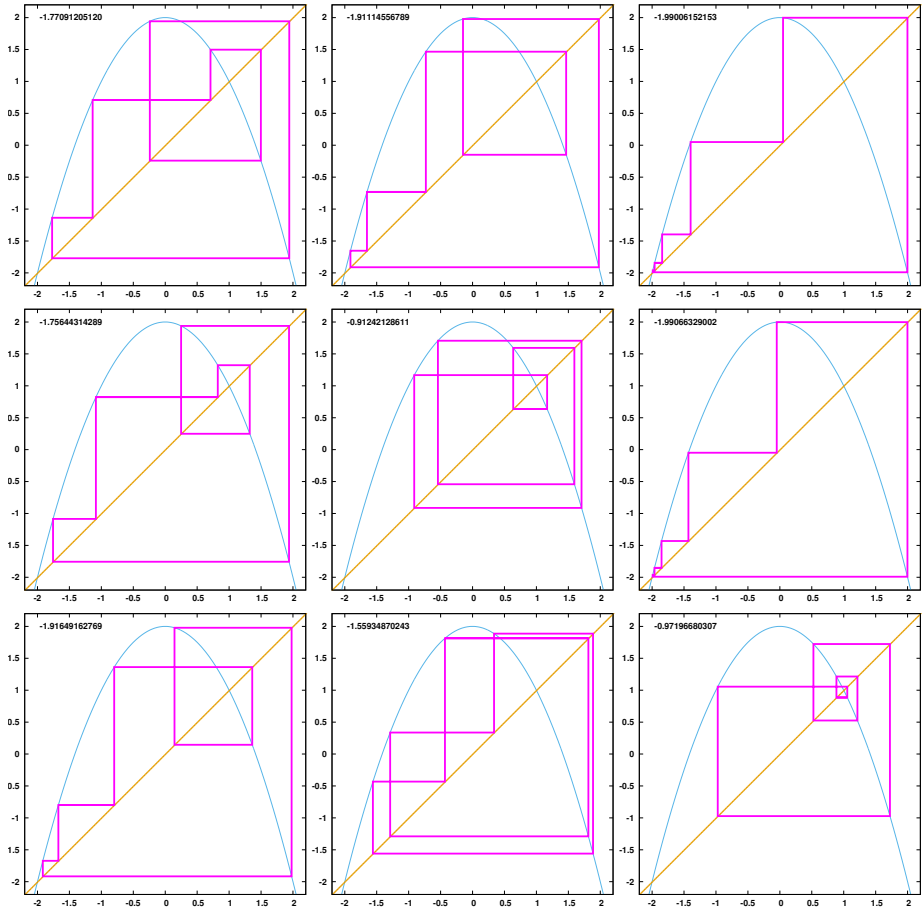}
\caption{\protect
  Return maps $x_t \times x_{t+1}$ for the nine period-6 orbits.
  Numbers refer to the leftmost orbital coordinate.
  Some of the orbits are topologically identical,
  despite their very distinct algebraic character.
}
\label{fig:fig01} 
\end{figure*}
}

The orbital points for all nine period-six orbits are collected in
Table \ref{tab:tab01}, together with the sums $\sigma_{6,j}$.
Return maps for all nine orbits are illustrated in Fig.~\ref{fig:fig01}.
Numbers inside panels identify the leftmost orbital point.
From Fig.~\ref{fig:fig01}  one sees that some orbits are
topologically identical despite the very distinct nature of the
algebraic numbers underlying them.

It is interesting to mention that Eqs.~(\ref{o61}) and (\ref{o62}) imply
a novel twist in the current understanding of polynomial interdependence.
As individual orbits, they are obtained as two  ``projections''
arising from a common mathematical origin, the carrier $\psi_6(x)$.
Therefore, rather than independent orbits, they are in a certain sense
a kind of ``conjugated'' orbits.
Furthermore, since all nine orbits arise from the same carrier $\psi_6(x)$,
the pair of orbits is also conjugated to the remaining seven orbits.
This illustrates the existence of a complex and subtle
{\sl arithmetical interdependence} lurking among such orbits,
apparently rather different from the usual field isomorphisms familiar
from Galois theory of equations. 
Orbital carriers allow conjugated orbits to have coefficients from
very distinct number fields, a concept alien to the standard theory.

As for the remaining polynomials,
$o_{6,1}(x)$ factors into a pair of cubics over $\mathbb Q(\sqrt{13})$.
The cluster $c_{6,2}(x)$ factors into two equations of degree nine
over $\mathbb Q(\sqrt{21})$.
However, these polynomials mix roots
   of distinct orbits, since their degree is not a multiple of six.
For $c_{6,1}(x)$, six cubics are obtained over $\mathbb Q(\sqrt[3]{\alpha})$,
where $\alpha=-154+42\sqrt{-3}-18\sqrt{-7}+30\sqrt{21}$.
For $c_{6,2}(x)$, eight cubics are obtained over $\mathbb Q(\sqrt{\beta})$,
where $\beta=65-13\sqrt{5} + 15\sqrt{15} - 3\sqrt{65}$.
These factorizations provide explicit and exact solutions for all period-six
orbits.
Note that the factors of $\mathbb S_k(\sigma)$ reveal how
orbits are distributed into clusters and single orbits, if any.

\section{Orbital inheritance}

A simple example allows one to grasp easily what inheritance
means\cite{epl,jg00,prep}.
To this end, we apply the nonlinear transformation $x^3 -3x$ to 
$o_{6,2}(x)$, obtaining the identity:
\begin{equation}
c_{6,1}(x) = o_{6,2}(x^3-3x).
\end{equation}
This identity shows that, as soon as the roots $z_i$ of $o_{6,2}(x)=0$
are determined, three new orbits follow from the zeros of the six cubics
\begin{equation}
  x^3 -3x -z_i=0.
\end{equation}
Therefore,
since $o_{6,2}(x)$ factors into a pair of cubics over $\mathbb Q(\sqrt{21})$,
{\small
\begin{eqnarray}
  o_{6,2}(x) &=& \big(x^3+\tfrac{1}{2}(1-\sqrt{21})x^2
  -\tfrac{1}{2}(1+\sqrt{21})x +\tfrac{1}{2}(5+\sqrt{21})\big)\times\cr
    && \big(x^3+\tfrac{1}{2}(1+\sqrt{21})x^2
        -\tfrac{1}{2}(1-\sqrt{21})x +\tfrac{1}{2}(5-\sqrt{21})\big), \label{cuc}
\end{eqnarray}}
their roots provide exact analytical solutions in terms of radicals
for all orbital points of the 18th-degree cluster $c_{6,1}(x)=0$.
Such exact solutions are simple cascades, towers, of
relative cubic irrationalities\cite{som07}.
Incidentally, Maple surprisingly fails to solve the sextic
$o_{6,2}(x)$ using 
\verb|aux:=solve(o62,x); convert(aux[1],radical);|
But it correctly breaks $o_{6,2}(x)$  into a pair of cubics when adding input
from the sextic discriminant: \verb|factor(o62,21^(1/2));|

What about the inherent character of the irrationalities underlying period-six
orbital point? This question is particularly interesting because, while
compositions of relative quadratic irrationalities are long known\cite{som07},
the considerably more complicated structures arising from nested cyclic cubic 
irrationalities remains essentially open\cite{df64}.
Thus, present day computer algebra systems still have to grapple with
difficulties to simplify expressions containing cubic and higher
roots\cite{la94,la03,bcz7,bcz}.
By way of illustration, consider to reassemble the cubics in Eq.~(\ref{cuc})
starting from the exact expressions of their three roots.
In this case, we get the leftmost
number below as the second coefficient in the topmost equation, not
its most simplified version:
{\small
\begin{equation}
  \frac{119-21\sqrt{-3}-27\sqrt{-7}+31\sqrt{21}}
       {-77+21\sqrt{-3}+9\sqrt{-7}-15\sqrt{21}}  =
 -\frac{49+11\sqrt{21}}{2(14+3\sqrt{21})}   =
       \tfrac{1}{2}(1-\sqrt{21}) \simeq -1.791287.
\end{equation}}
Similarly garbled expressions are obtained for all other coefficients
in Eq.~(\ref{cuc}).
The good news is that such expressions provide clues regarding the subfield
structure underlying the solutions. Clues may be also obtained from the
algebraic numbers solving the factors in $\mathbb S_k(\sigma)$.

{\small
\begin{table}[!bth]
\tbl{Data summary of polynomial factors as a function of the period $k$.
  Type refers either to orbits $o_{k,j}$ or orbital clusters $c_{k,j}$,
  $\partial$ is the degree of the corresponding
  polynomial, $D=\Delta$ are the standard polynomial and field discriminants,
  and $L$ is the length, the number of digits of the discriminants.
  For a given $k$, highlighted cells indicate discriminants arising from
  identical prime numbers (see text).}
{\begin{tabular}{@{}|c|c|c|c|c|  |c|c|c|c|c|@{}} 
\hline
$k$ & Type & $\partial$ & $D=\Delta$ & $L$ & 
$k$ & Type & $\partial$ & $D=\Delta$ & $L$ \\
\hline
3&\xxb{$\mathbf o_{3,1}$}& 3& $7^2$ & 2 & 9& \xxb{$\mathbf o_{9,1}$}& 9& $3^{22}$ & 11\\
 & \xxb{$\mathbf o_{3,2}$}& 3& $9^2$ & 2 &  & \xxb{$\mathbf o_{9,2}$}& 9& $19^8$  & 11\\
4& \xxb{$\mathbf o_{4,1}$}& 4 & $3^2\cdot 5^3$ & 4& & $c_{9,1}$& 18& \iw$3^9\cdot 19^{17}$ &  27\\
 & $c_{4,1}$ & 8 & $17^7$  & 9 &        &$c_{9,2}$ & 36&    $73^{35}$ &  66\\ 
5& \xxb{$\mathbf o_{5,1}$} & 5  & $11^4$ & 5    &  & $c_{9,3}$& 54& \iw$3^{81}\cdot 19^{51}$ & 104\\
 &$c_{5,2}$ & 10 & $3^5\cdot 11^9$ & 12 &  &$c_{9,4}$&162& \iw$3^{405}\cdot 19^{153}$ & 389 \\
 &$c_{5,3}$ & 15 & $31^{14}$ & 21 &       &$c_{9,5}$&216& $7^{180}\cdot 73^{213}$ & 550\\
\hline
6& \xxb{$\mathbf o_{6,1}$} & 6  & $13^5$ & 6
      &  10& \xxb{$\mathbf o_{10,1}$}& 10& $5^{17}$ & 12\\
& \xxb{$\mathbf o_{6,2}$} & 6  &\bl $3^3\cdot 7^5$ & 6
      &    & $c_{10,1}$& 20& $41^{19}$ & 31\\
& $c_{6,1}$ & 18 &\bl $3^{27}\cdot 7^{15}$ & 26
        & & $c_{10,2}$& 30& $3^{15}\cdot 31^{29}$ & 51\\
& $c_{6,2}$ & 24 & $5^{18}\cdot 13^{22}$ & 38
        &   & $c_{10,3}$& 80& \by$5^{60}\cdot 41^{78}$ & 168\\
7& $c_{7,1}$ & 21 & $43^{20}$ & 33 &   & $c_{10,4}$& 150& $11^{135}\cdot 31^{145}$ & 357\\
 & $c_{7,2}$ & 42 & $3^{21}\cdot 43^{41}$  &77 &   & $c_{10,5}$& 300& $3^{150}\cdot 11^{270}\cdot 31^{290}$ & 786\\
 & $c_{7,3}$ & 63 & $127^{62}$& 131 &   &  $c_{10,6}$& 400& \by$5^{700}\cdot 41^{390}$ & 1119\\
\hline  
8 & $c_{8,1}$ & 16 & $3^8\cdot 17^{15}$ & 23 &  11
   & \xxb{$\mathbf o_{11,1}$}& 11& $ 23^{10}$ & 14\\
& $c_{8,2}$ & 32 & $5^{24}\cdot 17^{30}$ & 54 &
&   $c_{11,1}$& 44& $89^{43}$ & 84\\
&  $c_{8,3}$ & 64 & $3^{32}\cdot 5^{48}\cdot 17^{60}$ & 123 &
   &  $c_{11,3}$& 341& $ 683^{340}$ & 964\\
& $c_{8,4}$ &128 & $257^{127}$ & 307 &   
 &  $c_{11,4}$& 682& $3^{341}\cdot 683^{681}$ & 2093\\
&&&&   &  &   $c_{11,5}$& 968& $23^{924}\cdot 89^{957}$ & 3124\\
\hline
\end{tabular}}\label{tab:tab02}
\end{table}}

\section{Inheritance systematics up to periods $k\leq 12$}

\subsection{Periods $k\leq 11$}

Using a slightly adapted version of the  {\it ad-hoc} Maple driver
given in appendix A,
we computed systematically all genuine factors defining orbits with
period $k\leq 12$. A summary of the relevant data obtained for
$k\leq 11$ is given in Table \ref{tab:tab02}.
This table reveals a number of interesting facts and trends:
\begin{enumerate}[noitemsep,topsep=0pt] 
\item The growth of the number of single orbits is much smaller than 
  cluster growth.
\item Periods $k=7$ and $k=8$ contain only orbital clusters, no single orbits.
\item Orbits and clusters are all {\sl monogenic}, i.e.~the discriminant $D$ of
  their minimal polynomial coincides with their field discriminant $\Delta$.
  Therefore,  orbits and clusters  admit power integral bases.
  For details, see Ref.~\cite{mono}. 
\item The degree of single orbits and clusters is always a multiple
  of the period $k$. 
\item As indicated by the length $L$ giving the number of digits in
  the discriminants,
  $D$ and $\Delta$ grow fast with the period. However, they contain powers
  of relatively small prime numbers.
\item The discriminants of, e.g., $c_{11,5}$ contain 3124 digits.
  It would be computationally hard to factor it if it was not a
  simple product of powers of a few identical and small primes, $23$ and $89$. 
\item The highlighted values of $D=\Delta$ for $k=6, 9$, and $10$ summarize
  all cases of inheritance found for $k\leq11$.
\item For $k=6$ the ratio of the polynomial degrees are
  $\partial(c_{6,1})/\partial(o_{6,1}) =3$. Similarly, for $k=9$ the ratios are
  $\partial(c_{9,4})/\partial(c_{9,3}) = \partial(c_{9,3})/\partial(c_{9,1})=3$.
  Inheritance among these orbits involves the aforementioned cubic
  transformation:
  $c_{9,3}(x) \equiv c_{9,1}(x^3-3x)$ and $c_{9,4}(x) \equiv c_{9,3}(x^3-3x)$.
\item In contrast, for $k=10$ the ratio is
  $\partial(c_{10,6})/\partial(o_{10,3}) =5$,
  implying inheritance involving a quintic nonlinear transformation\cite{jg00}.
  In this case, we have $c_{10,6}(x) \equiv o_{10,3}(x^5-5x^3+5x)$.
\item For a given period $k$, the discriminants $D$ and $\Delta$ involve
  certain combinations of a small set of primes. We were not able to find
  interconnections
  between orbits with discriminants arising from powers of distinct primes,
  although we see no reason to rule out the possibility of
  intricate interconnections yet to be discovered.
\item From Table \ref{tab:tab02}, it seems reasonable to conjecture inheritance
  to exist among polynomials with discriminants composed by powers of the
  same primes.
\end{enumerate} 
  
{\small  
\begin{table}[!tbh]
  \tbl{Individual factors of the 4020th degree
    polynomial containing all period twelve orbits and orbital clusters.
    Here, $\partial$ refers to the  degree of individual factors,
    while length is the number of digits contained in the
    discriminants $D=\Delta$.
    Similar highlighting is used for discriminants defined
    by identical prime numbers. No more than pairs of interdependent
    orbits are  observed.}
{\begin{tabular}{@{}|c|c|c|c|@{}} 
\hline    
      $\ell$ & Degree $\partial$ & $D=\Delta$ & Length \\
      \hline
1 & 12 & $5^9\cdot7^{10}$  & 15\\
2 & 12 & {$3^{18}\cdot5^9$}  & 15\\
3 & 12 & \iw$3^6\cdot13^{11}$ & 16\\
4 & 24 & \aw$3^{12}\cdot 5^{18} \cdot 7^{20}$ & 36\\
5 & 36 & \iw$3^{54}\cdot 13^{33}$ & 63\\
6 & 36 & $7^{30}\cdot13^{33}$  & 63\\
7 & 48 & \iy$3^{24}\cdot5^{36}\cdot 13^{44}$ & 86\\
8 & 72 & \aw$3^{108}\cdot 5^{54}\cdot 7^{60}$ & 140\\
9 & 72 & \bl$3^{36}\cdot 7^{60}\cdot 13^{66}$ & 142\\
10& 120& $241^{119}$ & 284\\
11& 144& $5^{108}\cdot 7^{120}\cdot 13^{132}$ & 324\\
12& 144& \iy$3^{216}\cdot 5^{108}\cdot 13^{132}$ & 326\\
13& 216& \bl$3^{324}\cdot 7^{180}\cdot 13^{198}$ & 528\\
14& 288& \am$3^{144}\cdot 5^{216}\cdot 7^{240}\cdot 13^{264}$ & 717\\
15& 864& \am$3^{1296}\cdot 5^{648}\cdot 7^{720}\cdot 13^{792}$& 2562\\
16& 1920& $17^{1800}\cdot 241^{1912}$ & 6770\\
\hline
\hline
\end{tabular}}\label{tab:tab03}
\end{table}}

\subsection{Period $k=12$}

Table \ref{tab:tab03} summarizes data obtained for the sixteen individual
factors resulting from the computation and factorization of the 4020th degree
polynomial which contains all genuine period twelve orbits and clusters.
These factors corroborate the properties listed above for $k\leq11$.
Note the fast increase in the number of digits of the discriminants,
which for $c_{12,16}(x)$ contains no less than 6770 digits.
In order to factor arbitrary numbers of this size, computers need to check
numbers of the order of the size of the square-root of the number
to be factored, in the present case roughly $10^{3385}$.

\begin{table}[!tbh]
  \tbl{\protect Growth of the number $N_k$ of periodic orbits, as a
    function of the
    period $k$. The  number of orbits roughly doubles as
    $k$ increases. For simple equations and Maple implementations
    to obtain arbitrary values of  $N_k$ see     Refs.~$^{27,28}$.}
{\begin{tabular}{@{}|c||c|c|c|c|c|c|c|c|c|@{}} 
\hline
$k$ &  12 &  13 & 14 & 15 & 16 & 17 & 18  &19 & 20\\
\hline
$N_k$         & 335 & 630 & 1161& 2182 & 4080 & 7710 & 14532 & 27594 & 52377\\
$N_k/N_{k-1}$& 1.80& 1.88& 1.84& 1.88 & 1.87 & 1.90 & 1.88 & 1.90 & 1.90\\
\hline
\end{tabular}}\label{tab:tab04}
\end{table}

Numbers with 6770 digits are well beyond the capabilities of factorization,
and also well beyond the numbers currently used in data encryption.
For instance, consider that the lifetime of the universe, currently
estimated to be some 13.8 billion years, roughly $10^{18}$ seconds,
a number with 19 digits.
Assuming a computer able to test one million factorizations per second,
during the lifetime of the universe it would be able to check some
$10^{24}$ possibilities.
However, for $6770$ digits, roughly $10^{6770}$,
one would need to check $10^{3385}$ possibilities, meaning that the
time to do this amounts to roughly $10^{3385-24}=10^{3361}$ times the lifetime
of the universe!
Fortunately, however, the very big numbers in Table \ref{tab:tab03}
involve products of just a few  and small primes, allowing them
to be factored, as indicated in the Table.
The passage here is exceedingly narrow.
Slight changes in the coefficients may preclude factorization.

The most conspicuous difference when comparing the numbers in
Table \ref{tab:tab03} with analogous results for the lower
periods in Table \ref{tab:tab02} is the surprising increase
of the number of polynomials
displaying inheritance. For instance, abbreviating $X=x^3-3x$,
we find the following five nonlinear interconnections among
polynomials of quite high degrees:
$c_{12,5}(x) \equiv o_{12,3}(X)$,
$c_{12,8}(x) \equiv c_{12,4}(X)$,
$c_{12,12}(x) \equiv c_{12,7}(X)$,
$c_{12,13}(x) \equiv c_{12,9}(X)$, and
$c_{12,15}(x) \equiv c_{12,14}(X)$.
The verification of these identities requires {\it ad-hoc} handling because
of recurring Maple warnings ``stack limit reached''.

Table \ref{tab:tab04} illustrates how fast the number of orbits grows as a
function of the period $k$.
A simple and explicit formula and its Maple implementation
to compute such growth is available in the literature\cite{jg07,count}.
It would be interesting to extend the present calculations and check
inheritance for the promising cases $k=14, 15$, $16$ and $18$, something
that should be feasible already by someone with access to more powerful
resources than available to us.

Two additional aspects are worth mentioning:
First, periodic orbits may be found by studying preperiodic points\cite{prep}.
Such procedure involves just straightforward but somewhat tedious computations,
due to the large number of factors and orbits involved.
Fortunately, the procedure involving preperiodic points may be programmed
to run automatically.
Second, by a process of reverse engineering and by suitably summing orbital
points, one may recover the several individual factors arising in the
$\mathbb S_k(\sigma)$ polynomials.
For instance, in Appendix B we compute explicitly the three factors composing
$\mathbb S_7(\sigma)$.
For single orbits the factors are very simple to find.
For instance, the single period-twelve orbits are
{\small
\begin{eqnarray}
  o_{12,1}(x) &=& {x}^{12}+{x}^{11}-12\,{x}^{10}-11{x}^{9}+54{x}^{8}+43{x}^{7}
       -113{x}^{6}-71{x}^{5}\cr
      &&\quad\ \; +110{x}^{4}+46{x}^{3}-40{x}^{2}-8x+1,\\
  o_{12,2}(x) &=&{x}^{12}-12{x}^{10}+{x}^{9}+54{x}^{8}-9{x}^{7}-112{x}^{6}
       +27{x}^{5}+105{x}^{4}\cr
      &&\quad\ \; -31{x}^{3}-36{x}^{2}+12x+1,\\
       o_{12,3}(x) &=&{x}^{12}+{x}^{11}-12{x}^{10}-12{x}^{9}+53{x}^{8}+53{x}^{7}
       -103{x}^{6}-103{x}^{5}\cr
  &&\quad\ \; +79{x}^{4}   +79{x}^{3}-12{x}^{2}-12x+1,
\end{eqnarray}}\noindent
and we immediately recognize that $s(s+1)^2$ are the linear factors of
$\mathbb S_{12}(\sigma)$, a curious degenerate multiplicity situation which
seems to foretell that $\psi_{12}(x)$ will be  a reducible polynomial.
Analogously, linear factors of  $\mathbb S_{k}(\sigma)$ may be
read directly from the coefficients of the orbits:
{\small
\begin{eqnarray}
  o_{9,1}(x) &=& x^9-9x^7+27x^5-30x^3+9x-1,\\
  o_{9,2}(x) &=& x^9-x^8-8x^7+7x^6+21x^5-15x^4-20x^3+10x^2+5x-1,\\
  o_{10,1}(x) &=& {x}^{10}-10{x}^{8}+35{x}^{6}-{x}^{5}
                -50{x}^{4}+5{x}^{3}+25{x}^{2}-5x-1,\\
  o_{11,1}(x) &=& {x}^{11}-{x}^{10}-10{x}^{9}+9{x}^{8}+36{x}^{7}-28{x}^{6}
                 -56{x}^{5}+35{x}^{4}+35{x}^{3}-15{x}^{2}-6\,x+1.
\end{eqnarray}}

It is quite challenging to decompose orbital clusters combining more than
two orbits, particularly those combining an odd number of orbits.
However, the coefficients of such decompositions hide the secretest
truth and most interesting relations among numbers which fix orbital
individuality.


\section{Conclusions and outlook}

This paper presented explicit expressions for orbital carriers of periods
$4$, $5$, and $6$. In addition, the systematics of orbital inheritance was
considered for all periods $k\leq12$. Evidence was found that inheritance
becomes more abundant as the period increases.
Useful insight was obtained from the exact properties of equations of motion,
instead of orbital points.
An interesting open challenge is to compute the distinct factors arising for
orbits of periods $k=14, 15$, $16$ and $18$, and to check if they also
display inheritance and relations with orbits of lower periods,
if any. A much harder problem seems to be to find out if orbits not displaying
inheritance may nevertheless display some other type of interdependence.
If found, this would certainly reveal unanticipated interconnections
among families of algebraic numbers.

As it is visible from Tables \ref{tab:tab02} and \ref{tab:tab03},
the growth of the polynomial degrees $\partial_k$ as a function of $k$
and their partition into proper divisors of $k$ are interesting
open combinatorial questions.
What is the mechanism behind the decomposition of the number $N_k$ of periodic
orbits into the several degrees $\partial_k$ of the polynomial set defining
$k-$periodic orbits?
For instance, the 4020th degree polynomial of period-12 orbits is partitioned
into sixteen factors recorded in Table \ref{tab:tab03}.
What would be, say, the corresponding partition for the 16254th degree polynomial
corresponding to period-14 orbits and clusters?
Or the 32730th degree polynomial for period-15?
Or the 65280th degree polynomial for period-16?
Note that the partitions listed in Table \ref{tab:tab02} are not unique:
for $k=6$, instead of $6+6+18+24$, we could equally well have 
$12+18+24$,  $12+12+30$, etc.
Such alternative partitions, however,
are never observed in the present context.
It is clear that the partition sets have many elements, and an interesting
combinatorial challenge is to count them all and to predict partitions
that may be observed for a given period of a given map.

Finally, for applications in physics and dynamical systems,
it is of interest to mention that in algebraic number theory one knows that
every cyclotomic field is an Abelian extension of the rational
numbers $\mathbb Q$.
In this context, an important discovery is the so-called Kronecker-Weber theorem,
stating that every finite Abelian extension of $\mathbb Q$ can be generated
by roots
of unity, i.e.~Abelian extensions are contained within some cyclotomic field.
Equivalently, every algebraic integer whose Galois group is Abelian
can be expressed as a sum of roots of unity with rational coefficients.
For details see, e.g., Edwards\cite{edw}.
The study of the partition generating limit of the quadratic map
$x_{t+1}=a-x_t^2$ seems to lend hope that for $a=2$ the map may
also share an analogous correspondence with Abelian equations as the one
embodied in the Kronecker-Weber theorem\cite{s98},
which is intrinsically related to the cyclotomic polynomials generated
by the map when $a=0$, whose dynamics, unbeknownst to him, was studied
by Gauss in {\it Sectio Septima} of his {\sl Disquisitiones Arithmetic\ae}.
Such enticing possibility of correspondence deserves to be
further investigated.


\section*{Acknowledgments}
This work was started during a visit to the Max-Planck Institute for
the Physics of Complex Systems, Dresden, gratefully supported by an
Advanced Study Group on {\sl Forecasting with Lyapunov vectors}.
The author was partially supported by CNPq, Brazil, grant 304719/2015-3.

\appendix

\section{Maple driver to generate period six orbits and clusters}

{\small
\begin{verbatim}
a := 2:
x[1]:= a - x*x:       x[2]:= a - x[1]*x[1]: x[3]:= a - x[2]*x[2]:   
x[4]:= a - x[3]*x[3]: x[5]:= a - x[4]*x[4]: x[6]:= a - x[5]*x[5]:  
aux := factor(x-x[6]);
##  FAKE period six orbits, containing repeated points:
 per1 := op(1,aux)*op(2,aux);    per2 := op(3,aux); 
 per3 := op(4,aux)*op(5,aux);
##  GENUINE period six orbits, containing NO repeated points:
o61:= op(7,aux); o62:= op(6,aux); c61:= op(9,aux); c62:= op(8,aux); 
\end{verbatim}}
The above assignments are correct under Maple 2014,
but are easy to adjust if emerging differently.
Manifestly, the driver above may be easily adapted to generate equations
for other periods.

\vspace{-0.5truecm}

\section{Determination of the three factors composing $\mathbb S_7(\sigma)$}

Here, in contrast to the arithmetic work done so far, we resort to
numerically computed orbital points to illustrate how to find
exact representations for the individual factors composing
$\mathbb S_7(\sigma)$. 
The three clusters whose roots give all period-seven orbital points
may be easily generated by slightly adapting the Maple driver given in
Appendix A. Such clusters read as follows:
{\small
\begin{eqnarray}
c_{7,1}(x) &=&   {x}^{21}-{x}^{20}-20\,{x}^{19}+19\,{x}^{18}+171\,{x}^{17}
                -153\,{x}^{16}-816\,{x}^{15}+680\,{x}^{14}\cr
 && \quad\quad +2380\,{x}^{13}-1820\,{x}^{12}-4368\,{x}^{11}+3003\,{x}^{10}
                +5005\,{x}^{9}-3003\,{x}^{8}\cr
 && \quad\quad -3432\,{x}^{7} +1716\,{x}^{6}+1287\,{x}^{5}-495\,{x}^{4}
                -220\,{x}^{3}+55\,{x}^{2}+11\,x-1,\\
c_{7,2}(x) &=& {x}^{42}+{x}^{41}-42\,{x}^{40}-42\,{x}^{39}+  \cdots
           -3267\,{x}^{4}-3267\,{x}^{3}+44\,{x}^{2}+44\,x+1,\\    
c_{7,3}(x) &=&{x}^{63}-{x}^{62}-62\,{x}^{61}+61\,{x}^{60}  +\cdots
             +40920\,{x}^{4}+5456\,{x}^{3}-496\,{x}^{2}-32\,x+1.
\end{eqnarray}}%
From them, we extract the $(21+42+63)/7=18$ orbits
summarized in Table \ref{tab:tab0a}.
After rounding off the real coefficients in the products below,
we easily get the exact representations of the three factors
composing $\mathbb S_7(\sigma)$, all with degree multiple of three:
{\small
\begin{eqnarray}
  \prod_{j=1}^3 (\sigma-\sigma_{7,j}) &=& {\sigma}^{3}-{\sigma}^{2}-14\sigma-8,\\
  \prod_{j=4}^9 (\sigma-\sigma_{7,j}) &=& {\sigma}^{6}+{\sigma}^{5}-39\,{\sigma}^{4}
                    +63\,{\sigma}^{3}+110\,{\sigma}^{2}-136\sigma-128,\\
  \prod_{j=10}^{18} (\sigma-\sigma_{7,j}) &=&{\sigma}^{9}-{\sigma}^{8}-56{\sigma}^{7}
                   +118{\sigma}^{6}+573{\sigma}^{5}-1249{\sigma}^{4}
          -1582{\sigma}^{3}+2700{\sigma}^{2}+1576\sigma-32.\nonumber
\end{eqnarray}}%

{\small
\begin{table}[!bht]
  \tbl{The eighteen period-seven orbits, characterized by one orbital point
    and the sum $\sigma_{7,j}$ of all points. The remaining
      orbital points follow by iterating $x_{t+1}=2-x_t^2$.}
{\begin{tabular}{@{}|c||c|c|@{}} 
\hline
Orbit &  $x_1$ &  $\sigma_{7,j}$ \\
\hline
$o_{7,1}$ & -1.97868673615022039558470712622 & -2.88823600884341144649527347953\\
$o_{7,2}$ & -1.81089647498629323144358511206 & -0.61507162581156506493032243917\\
$o_{7,3}$ & -1.04188068097586057235256289907 &  4.50330763465497651142559591867\\
\hline
$o_{7,4}$ & -1.99762811048164609235032811636 & -7.24813219626988235042435866316\\
$o_{7,5}$ & -1.88487616566742883844738056630 & -1.22906022702843317616182868886\\
$o_{7,6}$ & -1.94098358831481064644663464031 & -0.774908002370389501560130448853\\
$o_{7,7}$ & -1.61228898351072554484668557445 &  1.84413185283999824109215112803\\
$o_{7,8}$ & -1.71974598668362014848475373830 &  2.74482456161490583899876274448\\
$o_{7,9}$ & -1.20298163000374078956875931348 &  3.66314401121380094805540392837\\
\hline
$o_{7,10}$ & -1.99755283242852256525640526121 & -7.17543506383968793122213507689\\
$o_{7,11}$ & -1.97801140897626144475812107452 & -2.93599431271533079530605723303\\
$o_{7,12}$ & -1.88125825920768775450635179933 & -1.60237082080316266127006810916\\
$o_{7,13}$ & -1.93911972959649314295081201674 & -0.525572883742883886011679753072\\
$o_{7,14}$ & -1.80499303815485256128035783817 &  0.0196507480462068262052865602836\\
$o_{7,15}$ & -1.60040839696003410107405022267 &  2.19795804752238429773060428492\\
$o_{7,16}$ & -1.71107014481703192827696436581 &  2.36473479389711276652833789748\\
$o_{7,17}$ & -1.17956942634103896113267847206 &  3.29077832666324426013918210422\\
$o_{7,18}$ & -1.01424772773954618184418426842 &  5.36625116497211712320652932523\\
\hline
\end{tabular}}\label{tab:tab0a}
\end{table}
}

\noindent
Even though an expression for the period-seven carrier pair is still unknown,
we were nevertheless able to extract $\mathbb S_7(\sigma)$.
Its factors corroborate the three aggregates $c_{7,m}(x)$, $m=1,2,3$
and identify the
relative algebraic nature of the coefficients of individual orbits.
Manifestly,  the above procedure is valid generically and may
be applied to higher periods.
Separation of orbits into three groups in Table \ref{tab:tab0a}
was only possible due to the {\it a priori} knowledge of the three
``brute-force factors''
in Table \ref{tab:tab02} and given explicitly above, in $c_{7,m}(x)$.
However, using preperiodic points generated by an infinite family
$Q_\ell(x)$ of polynomials\cite{prep}, the same three groups my be discovered
independently, directly from numerically approximated orbital equations.
How to accomplish this will be presented in a forthcoming publication.




\end{document}